\documentclass[aps,floats,floatfix,showpacs,preprintnumbers,amssymb,prl,twocolumn,superscriptaddress,nofootinbib,nolongbibliography,reprint]{revtex4-1}

\usepackage{amssymb,amsmath,verbatim,mathtools,needspace,enumitem,etoolbox,graphicx,physics,microtype,afterpage,xspace,tabularx,lmodern,multirow,boldline, makecell, booktabs}

\usepackage{appendix}
\usepackage{tensor}
\usepackage{array}
\usepackage{siunitx}
\usepackage[normalem]{ulem}
\usepackage[dvipsnames, usenames]{xcolor}
\usepackage{xr-hyper}
\definecolor{linkcolor}{rgb}{0.0,0.3,0.5}
\usepackage[unicode, colorlinks=true, linkcolor=linkcolor, citecolor=linkcolor, filecolor=linkcolor, urlcolor=linkcolor, linktocpage, breaklinks]{hyperref}
\usepackage[all]{hypcap}
\usepackage[T1]{fontenc}
\usepackage[utf8]{inputenc}
\definecolor{cornellGreen}{HTML}{6EB43F}
\definecolor{cornellRed}{HTML}{B31B1B}
\hypersetup{colorlinks=true,citecolor=romared,linkcolor=romared,urlcolor=romared}

\definecolor{romared}{RGB}{142,0,28}

\newcommand{\beq}{\begin{eqnarray}}
\newcommand{\eeq}{\end{eqnarray}}

\usepackage{soul}

\newcolumntype{Y}{>{\centering\arraybackslash}X}

 \definecolor{mypurple}{RGB}{130, 0, 130}

\begin{document}
\title{
Resonances as Signatures of Scalar Clouds in Eccentric Extreme-Mass-Ratio Inspirals
}
\author{Qi-Xuan~Xu}
\email{qixuan.xu@tecnico.ulisboa.pt}
\affiliation{CENTRA, Departamento de Física, Instituto Superior Técnico – IST,\\
Universidade de Lisboa – UL, Avenida Rovisco Pais 1, 1049-001 Lisboa, Portugal}

\author{Richard~Brito}
\affiliation{CENTRA, Departamento de Física, Instituto Superior Técnico – IST,\\
Universidade de Lisboa – UL, Avenida Rovisco Pais 1, 1049-001 Lisboa, Portugal}

\author{Riccardo~Della~Monica}
\affiliation{CENTRA, Departamento de Física, Instituto Superior Técnico – IST,\\
Universidade de Lisboa – UL, Avenida Rovisco Pais 1, 1049-001 Lisboa, Portugal}

\author{Rodrigo~Vicente}
\affiliation{Gravitation Astroparticle Physics Amsterdam (GRAPPA),\\
University of Amsterdam, 1098 XH Amsterdam, The Netherlands}

\author{Chen~Yuan}
\affiliation{CENTRA, Departamento de Física, Instituto Superior Técnico – IST,\\
Universidade de Lisboa – UL, Avenida Rovisco Pais 1, 1049-001 Lisboa, Portugal}
\date{\today}
\begin{abstract}
Ultralight scalars arise naturally in many extensions to the Standard Model and are compelling dark matter candidates. 
Around spinning black holes, dense scalar clouds could form through the conversion of rotational energy into particles via black hole superradiance.
Extreme-mass-ratio inspirals (EMRIs) targeted by future space-based detectors will give us unparalleled access to the environments of massive black holes, allowing us to probe the presence of scalar clouds.
We consider EMRIs around a Schwarzschild black hole and show that \emph{eccentricity} induces a dense sequence of resonances in the scalar fluxes near the last stable orbit. These resonances arise only in a fully relativistic treatment, as they are intrinsically tied to the splitting between the azimuthal and radial orbital frequencies in the strong-field regime. By evolving the orbits adiabatically, we show that the resulting resonant transitions substantially enhance the exchange of energy and angular momentum between the EMRI and the scalar cloud, significantly amplifying the accumulated dephasing in the gravitational waveform relative to circular motion. Our results highlight the importance of eccentricity in shaping the observational signatures of EMRIs embedded in scalar clouds.
\end{abstract}
\maketitle
\noindent{\bf{\em Introduction.}}
Gravitational waves (GWs) provide a unique probe of compact-object mergers. While the LIGO--Virgo--KAGRA collaboration has now observed hundreds of such events~\cite{LIGOScientific:2025slb,LIGOScientific:2025brd,LIGOScientific:2025rid,LIGOScientific:2025rsn}, upcoming detectors will greatly enhance the precision of GW astronomy. Future space-based detectors such as the Laser Interferometer Space Antenna (LISA)~\cite{LISA:2024hlh}, TianQin~\cite{TianQin:2015yph}, and Taiji~\cite{Luo:2021qji} will open the milliHertz window where extreme-mass-ratio inspirals (EMRIs) are among the most promising sources. EMRIs consist of a stellar-mass compact object inspiraling into a massive black hole (MBH), accumulating tens to hundreds of thousands of cycles in the strong-field regime. This makes them unparalleled probes of MBH spacetimes and their astrophysical environments~\cite{Barausse:2014tra,Cardenas-Avendano:2024mqp,LISA:2022kgy,Vicente:2025gsg}.

A compelling scenario is the presence of an ultralight bosonic cloud around the primary black hole (BH), formed via superradiant instabilities~\cite{Brito:2015oca}. Such environments can leave a detectable imprint on the GW signal emitted by an EMRI~\cite{Cole:2022yzw,Khalvati:2024tzz}, potentially revealing the existence of new fundamental bosons. 
Environmental effects of boson clouds in EMRIs have been extensively studied using Newtonian approximations~\cite{Baumann:2018vus,Baumann:2019ztm,Baumann:2022pkl,Tomaselli:2023ysb,Boskovic:2024fga,Boskovic:2025ixx}, which are insufficient to capture the dynamics in the late inspiral phase, while relativistic calculations have only considered circular, equatorial EMRIs in scalar clouds~\cite{Brito:2023pyl,Duque:2023seg,Dyson:2025dlj,Li:2025ffh,Keijzer:2026vul}.
However, realistic EMRIs are expected to possess non-vanishing eccentricity~\cite{Mancieri:2025cmx}. 

In this {\it Letter}, we reveal a new class of signatures for EMRIs embedded in scalar clouds, within the fully relativistic framework developed in our companion paper~\cite{Xu:2026aic}. We show that eccentricity triggers a dense series of resonant transitions between cloud states in the strong-field regime, amplifying the imprint of the cloud's presence in EMRI signals. These resonances arise from the split between azimuthal and
radial frequencies in the strong-field regime, a feature that is absent in Newtonian approximations.
Related resonant phenomena were first identified as a mechanism for the \textit{floating} or \textit{sinking} of circular inspirals in scalar-tensor theories~\cite{Cardoso:2011xi}, and were subsequently shown to be excited across a discrete spectrum of harmonics in eccentric systems~\cite{Fujita:2016yav}. In those works, the resonances were sourced by an intrinsic scalar charge. Here we show that similar resonances can arise solely from the gravitational interaction between a standard relativistic perturber and a scalar cloud environment.
While previous studies~\cite{Baumann:2018vus,Baumann:2019ztm,Berti:2019wnn,Boskovic:2024fga, Tomaselli:2024bdd, Tomaselli:2025jfo, Boskovic:2025ixx} also identified resonances in scalar clouds and discussed the associated orbital backreaction, their analyses were primarily Newtonian and focused on the early stages of the inspiral.
\\

\noindent{\bf{\em Framework.}} 
We consider a free complex scalar field $\mathbf{\Phi}$ minimally coupled to gravity in the presence of a point-like perturber with mass $m_p$, described by the action (we use geometrized units $G=c=1$)
\begin{equation}
S=\int{\dd}^{4}x\sqrt{-\mathbf{g}}\left(\frac{\mathbf{R}}{16\pi}-\mathfrak{L}[\mathbf{\Phi}]\right)-m_{p}\int_{\gamma}{\dd}\tau,
\end{equation}
where $\mathfrak{L}[\mathbf{\Phi}]=\partial_{\mu} \mathbf{\Phi}\partial^{\mu} \mathbf{\Phi}^*+\mu^2|\mathbf{\Phi}|^2$ is the scalar field's Lagrangian density and $\hbar \mu$ its mass, and $\gamma$ is the point-particle's worldline parametrized by its proper time $\tau$.
Varying the action with respect to the metric and the scalar field, one gets the Einstein-Klein-Gordon (EKG) field equations,
\begin{equation}
\begin{aligned}\label{eq: EKG field equation}
    \mathbf{G}_{\mu\nu} = 8\pi(T^{\mathbf{\Phi}}_{\mu\nu}+T^p_{\mu\nu}),\qquad
    \Box \mathbf{\Phi} = \mu^2 \mathbf{\Phi},
\end{aligned}
\end{equation}
where $\mathbf{G}_{\mu\nu}:= \mathbf{R}_{\mu\nu}-\mathbf{g}_{\mu\nu}\mathbf{R}/2$ and $\Box:=\mathbf{g}_{\mu\nu}\nabla^\mu\nabla^{\nu}$.  
The stress-energy tensor of the scalar field, $T_{\mu\nu}^\mathbf{\Phi}$, reads
\begin{equation}\label{eq: stress-energy tensor definition}
T_{\mu\nu}^\mathbf{\Phi}=2\partial_{(\mu}\mathbf{\Phi}\partial_{\nu)}\mathbf{\Phi}^*-\mathbf{g}_{\mu\nu}\left(\partial_\alpha\mathbf{\Phi}\partial^\alpha\mathbf{\Phi}^*+\mu^2|\mathbf{\Phi}|^2\right),
\end{equation}
whereas the stress-energy tensor of the point-particle (describing a small compact object), $T_{\mu\nu}^p$, is given by
\begin{equation}\label{eq: stress-energy tensor of the particle}
    T^{p}_{\mu\nu}=m_p\int_{\gamma} u_\mu u_\nu\frac{\delta^{(4)}[x^\mu-\mathbf{x}_p^\mu(\tau)]}{\sqrt{-\mathbf{g}}}\dd\tau,
 \end{equation}
with $u^{\mu}=d\mathbf{x}_p^{\mu}/d\tau$ the particle's four-velocity.
This theory has a global $U(1)$ symmetry with associated conserved Noether current and charge, respectively,
\begin{equation}\label{eq: current}
j^\mu=-i\left(\mathbf{\Phi}^*\partial^\mu\mathbf{\Phi}-\mathbf{\Phi}\partial^\mu\mathbf{\Phi}^*\right),\quad Q_\Sigma = \int_\Sigma \dd\Sigma_\mu \,j^\mu,
\end{equation}
where $\Sigma$ is a spacelike hypersurface.

We model the system perturbatively through a two-parameter expansion controlled by the small mass-ratio $q \equiv m_p/M \ll 1$ (with $M$ the MBH mass) and the background scalar field amplitude $\epsilon \ll 1$. The metric and scalar field are expanded as~\cite{Brito:2023pyl, Dyson:2025dlj}
\begin{equation}
    \mathbf{g}_{\mu\nu} = g_{\mu\nu} + \sum_{i,j} \epsilon^i q^j h_{\mu\nu}^{(i,j)},\quad \mathbf{\Phi} = \sum_{i,j} \epsilon^i q^j \phi^{(i,j)},
\end{equation}
where here $g_{\mu\nu}$ is the Schwarzschild background $\dd s^2 = -f \dd t^2 + f^{-1} \dd r^2 + r^2 \dd \Omega^2$ with $f=1-2M/r$. Plugging these expansions into the EKG equations yields a hierarchy of equations, which are solved order-by-order.

At order $\mathcal{O}(\epsilon^1 q^0)$ the metric is just Schwarzschild and the scalar field satisfies the 
homogeneous 
Klein-Gordon (KG) equation
\begin{equation}
    (\square^{(0)} - \mu^2)\phi^{(1,0)} = 0\,,
\end{equation}
which admits quasi-bound state solutions. Exploiting spherical symmetry, we decompose the field as $\phi^{(1,0)} = R^{b}_{n_i\ell_i}(r)Y_{\ell_im_i}(\theta, \varphi)e^{-i\omega t}$ with indices $\{n_i, \ell_i, m_i\}$ characterizing the solutions. Here, $Y_{\ell_im_i}$ are scalar spherical harmonics.
Quasi-bound state solutions with radial function $R^{b}_{n_i\ell_i}$ and their corresponding complex eigenfrequencies $\omega:=\omega_{n_i\ell_i}$ are found by imposing exponential decay at spatial infinity and purely ingoing waves at the event horizon and using the continued-fraction method developed in Refs.~\cite{Cardoso:2005vk, Dolan:2007mj}. 
Their total mass is $M_b=\omega\,  Q[\epsilon\phi^{(1,0)},g]$; for concreteness, we fix the arbitrary amplitude of $\phi^{(1,0)}$ such that $\epsilon\equiv \sqrt{M_b/M}$.

While superradiant growth of quasi-bound states requires a spinning BH~\cite{Ternov:1978gq,Detweiler:1980uk, Dolan:2007mj, Baumann:2019eav, Bao:2022hew}, for light scalars ($M\mu \ll 1$) superradiant extraction drives the BH toward a small saturation spin, and the scalar cloud is localized far from the horizon, where spin effects are subleading. Therefore, the Schwarzschild metric provides a good approximation for the cloud profile and associated radiation fluxes~\cite{Brito:2023pyl,Dyson:2025dlj}.

At order $\mathcal{O}(\epsilon^0 q^1)$ the metric perturbation $h_{\mu\nu}^{(0,1)}$ is governed by the linearized Einstein equations 
\begin{equation}
    \delta \mathbf{G}_{\mu\nu}[h^{(0,1)}] = 8\pi T_{\mu\nu}^{p}[g].    
\end{equation}
We solve these equations within the Regge-Wheeler-Zerilli formalism, with metric reconstruction 
performed according to the procedure in Refs.~\cite{Hopper:2010uv,Hopper:2015icj}. Metric perturbations are sourced by a small compact object following stable bound timelike geodesics in the Schwarzschild geometry.

In the equatorial plane, geodesics are determined by the dimensionless semi-latus rectum $p$ and eccentricity $e$, which are related to the conserved specific energy $\mathcal{E}$ and angular momentum $\mathcal{L}$ via~\cite{Cutler:1994pb,Barack:2010tm}
\begin{equation}
\mathcal{E}^2=\frac{(p-2-2e)(p-2+2e)}{p(p-3-e^2)},\quad\mathcal{L}^2=\frac{p^2M^2}{p-3-e^2}.
\end{equation}
The requirement of a stable bound orbit implies $p \geq p_{\rm LSO}\equiv 6+2e$,
with the equality at the last stable orbit~(LSO); 
for smaller values the orbit is unstable and the particle plunges. The source's eccentric motion restricts the metric perturbation to a discrete frequency spectrum
\begin{equation}
    \sigma_{mn} = m\,\Omega_\varphi + n\,\Omega_r, \qquad m,n\in\mathbb{Z},
\end{equation}
with azimuthal and radial orbital frequencies, respectively, $\Omega_\varphi(p,e)$ and $\Omega_r(p,e)$~\cite{Cutler:1994pb}.

The influence of the point particle on the scalar field first appears at $\mathcal{O}(\epsilon^1 q^1)$ through the inhomogeneous KG equation,
\begin{equation}\label{eq:11_KG}
    (\square^{(0)} - \mu^2)\phi^{(1,1)} = \nabla_{\mu}^{(0)}\bar{h}_{(0,1)}^{\mu\nu}\partial_{\nu}\phi^{(1,0)} + h_{(0,1)}^{\mu\nu}\nabla_{\mu}^{(0)}\partial_{\nu}\phi^{(1,0)},
\end{equation}
where $\bar{h}_{\mu\nu}^{(0,1)} = h_{\mu\nu}^{(0,1)} - \frac{1}{2}g_{\mu\nu}h^{(0,1)}$ is the trace-reversed metric perturbation. We solve for the scalar perturbation using a Green's function method, imposing outgoing boundary conditions at spatial infinity and ingoing ones at the MBH horizon; further details are given in~\cite{Xu:2026aic}.
\\

\noindent{\bf{\em Energy-momentum exchange and resonances.}}
The perturbative framework just described implies that, in addition to the usual GW fluxes, the EMRI exchanges energy and angular momentum with the ambient scalar field through gravitational scattering. In a steady-state regime, this exchange is encoded in the fluxes of energy and angular momentum carried by the scalar perturbations to spatial infinity and through the MBH horizon~\cite{Clough:2021qlv, Croft:2022gks, Annulli:2020lyc}. 

The orbit-averaged scalar fluxes are nonvanishing at order $\mathcal{O}(\epsilon^{2}q^{2})$, since they arise from quadratic terms in $\phi^{(1,1)}$.  
The energy fluxes are defined with respect to the stationary Killing vector $\xi^\mu_{(t)}=(\partial_t)^\mu$ as~\cite{Brito:2023pyl, Dyson:2025dlj}
\begin{subequations}\label{eq:dotE_Phi}
\begin{align}
    \dot E^{\Phi,\infty} &= -\lim_{r\to +\infty} r^2\int \dd\Omega\, T^{\phi^{(1,1)}}_{\mu r}\xi_{(t)}^{\mu}\,,\\ 
    \dot E^{\Phi,H} &= \lim_{r\to 2M} 4M^2 \int \dd\Omega\, T^{\phi^{(1,1)}}_{\mu t}\xi_{(t)}^{\mu}\,.
\end{align}
\end{subequations}
The angular momentum fluxes $\dot L^{\Phi,\infty/H}$ are obtained analogously by replacing $\xi_{(t)}^\mu$ with the axial Killing vector $\xi_{(\varphi)}^\mu=(\partial_\varphi)^\mu$. 
The charge evolves due to analogous fluxes to infinity and through the horizon, $\dot Q^{\infty/H}$, which are also order $\mathcal{O}(\epsilon^{2}q^{2})$. These induce changes in the cloud mass and angular momentum as $\dot M_b = \omega\, \dot Q$ and $\dot J_b = m_i\, \dot{Q}$.

\begin{figure}
    \centering
    \includegraphics[width=1.0\linewidth]{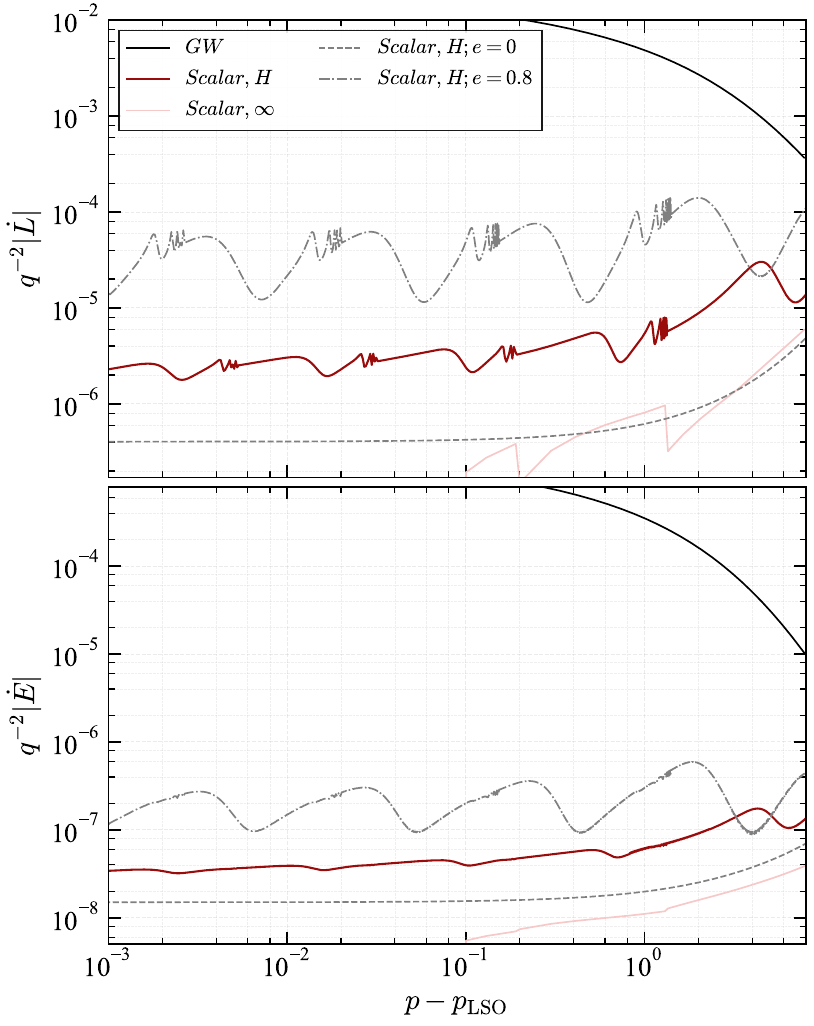}
    \caption{Angular momentum and energy loss rates $\dot{L}$ (\textit{top panel}) and $\dot{E}$ (\textit{bottom panel}) via gravitational radiation (black solid line) and scalar field scattering (red lines) shown as functions of semi-latus rectum $p$ for eccentricity $e=0.6$ (solid lines). For comparison, $\dot L^{s,H}$ and $\dot E^{s,H}$ for $e=0$ and $0.8$ (grey lines) are also shown. The background scalar field is in the prograde cloud state $\{n_i,\ell_i, m_i\} = \{0,1,1\}$ with $M\mu =0.2 $ and cloud mass $M_{b}/M = 0.05$. Resonances driven by eccentric motion in the strong field appear prominently in the torque from the scalar field.   
    Over the range of $p$ shown here, $\dot L^{s,H}$ and $\dot E^{s,H}$ are negative, whereas all other loss rates are positive.}
    \label{fig:Fluxes}
\end{figure}

Assuming that the inspiral proceeds on a timescale much longer than that to reach a steady-state energy-momentum exchange (i.e., using the adiabatic approximation), the point-particle's specific energy $\mathcal{E}$ and angular momentum $\mathcal{L}$ evolve according to:
\begin{equation}
\begin{aligned}\label{eq:balance_laws}
    &\dot{\mathcal{E}}=-m_p^{-1} \bigl( \dot{E}^{g,\infty}+\dot{E}^{g,H}+\dot{E}^{s,\infty}+\dot{E}^{s,H}\bigr),\\
    &\dot{\mathcal{L}}=-m_p^{-1} \bigl(\dot{L}^{g,\infty}+\dot{L}^{g,H}+\dot{L}^{s,\infty}+\dot{L}^{s,H}\bigr),
\end{aligned}
\end{equation}
where $\dot{E}^{g,\infty/H}$ and $\dot{L}^{g,\infty/H}$ are the GW energy and angular momentum fluxes at infinity and at the horizon, and we defined $\dot{E}^{s,\infty/H}\:=\:\dot{E}^{\Phi,\infty/H}+\omega\:\dot{Q}^{\infty/H}$ and $\dot{L}^{s,\infty/H}\:=\:\dot{L}^{\Phi,\infty/H}+m_i\:\dot{Q}^{\infty/H}$. These dissipation rates are computed in full generality (including eccentricity \textit{and} orbital inclination) for a Schwarzschild background in the companion paper~\cite{Xu:2026aic}. Here we focus on equatorial, prograde orbits. We neglect the conservative effects of the cloud's self-gravity on $\mathcal{E}$ and $\mathcal{L}$, which are non-vanishing at $\mathcal{O}(\epsilon^2)$, and $\mathcal{O}(\epsilon^2q^2)$ corrections to the GW fluxes~\cite{Brito:2023pyl,Keijzer:2026vul}. While these terms have not been computed yet for dipolar clouds, they are expected to be approximately degenerate with a redshift at small $M\mu$ (low compactness) or with a mass shift at large $M\mu$ (high compactness)~\cite{Duque:2023seg,Keijzer:2026vul}. Their relative importance at intermediate $M\mu$ for dipolar clouds is unclear. 
Conservative corrections from the cloud's self-gravity can shift both the orbital frequencies and the scalar energy levels, and therefore modify the locations and properties of the resonances~\cite{Kim:2025wwj}. However, the dominant resonances identified here involve Bohr-level transitions, for which self-gravity corrections to the level splitting are parametrically suppressed by a factor $(M\mu)^2$~\cite{Kim:2025wwj}. We therefore expect the dominant resonance structure to persist, although a self-consistent inclusion of conservative effects is required to determine accurate resonance locations and waveform dephasings. Eq.~\eqref{eq:balance_laws} also assumes that the orbit can be evolved only from radiative fluxes in the steady-state regime. An extension to this picture is to include non-radiative (local) torques using the torque-balance law of Ref.~\cite{Dyson:2026ddd}.

Here, our emphasis is on the role of resonances as characteristic signatures of scalar clouds in the signals of EMRIs. These resonances occur at an infinite set of orbital frequencies (that we label with $n_f$) given by
\begin{equation}\label{eq:resonances}
    \sigma^{n_f}_{mn}=\Re(\omega_{n_f \ell_f}) - \Re(\omega_{n_i\ell_i})\,,
\end{equation}
where we recall that $\omega_{n \ell}$ corresponds to an eigenfrequency of the KG equation with quantum numbers $\{n,\ell\}$ (note that in a Kerr background the eigenfrequencies would also depend on $m$). Eccentric EMRIs can generically excite such resonances in the strong-field regime, where space-based detectors will probe these systems.

Fig.~\ref{fig:Fluxes} compares the steady-state dissipation rates from gravitational radiation and from scalar-field scattering as a function of the semi-latus rectum, for an EMRI with $e=0.6$ in a prograde dipolar cloud $\{n_i,\ell_i,m_i\}=\{0,1,1\}$ with mass $M_b/M=0.05$ and coupling $M\mu=0.2$ (for an MBH of mass $M=10^6\,M_\odot$, this corresponds to bosons of mass $\hbar\mu\sim 3\times 10^{-17}\,\mathrm{eV}$). For comparison, we also show the horizon loss rates for circular geodesics ($e=0$) and for $e=0.8$. The total GW energy and angular-momentum fluxes are computed with the \texttt{FastEMRIWaveforms} package~\cite{chapman_bird_2025_15630565,Chua:2020stf,Katz:2021yft}. The scalar loss rates are dominated by the horizon terms $\dot L^{s,H}$ and $\dot E^{s,H}$, as in the circular case~\cite{Brito:2023pyl,Dyson:2025dlj,Li:2025ffh}. The step-like behavior seen in $\dot L^{s,\infty}$ is also expected and was first explained within Newtonian approximations~\cite{Baumann:2021fkf,Tomaselli:2023ysb}.  

\begin{figure*}
    \centering
    \includegraphics[width=0.48\textwidth]{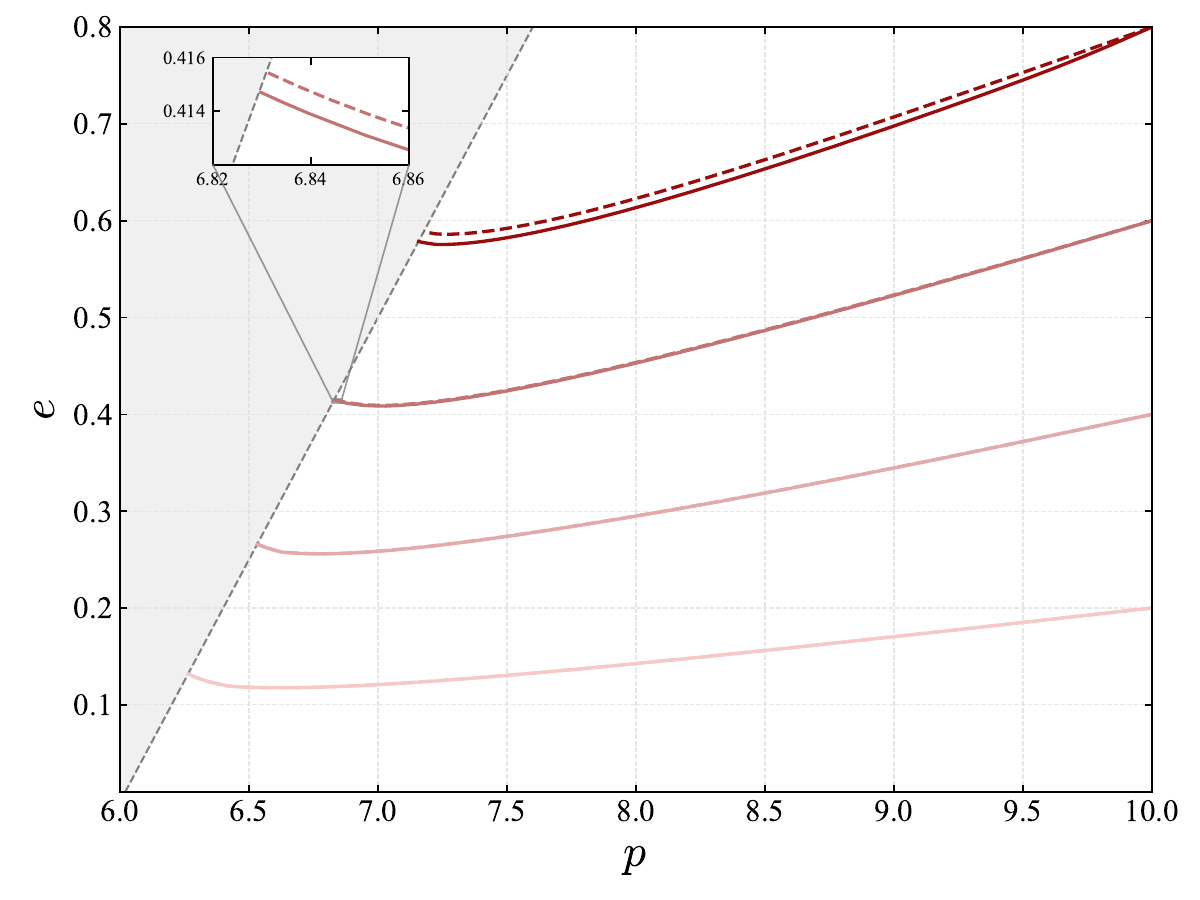}
    \includegraphics[width=0.48\textwidth]{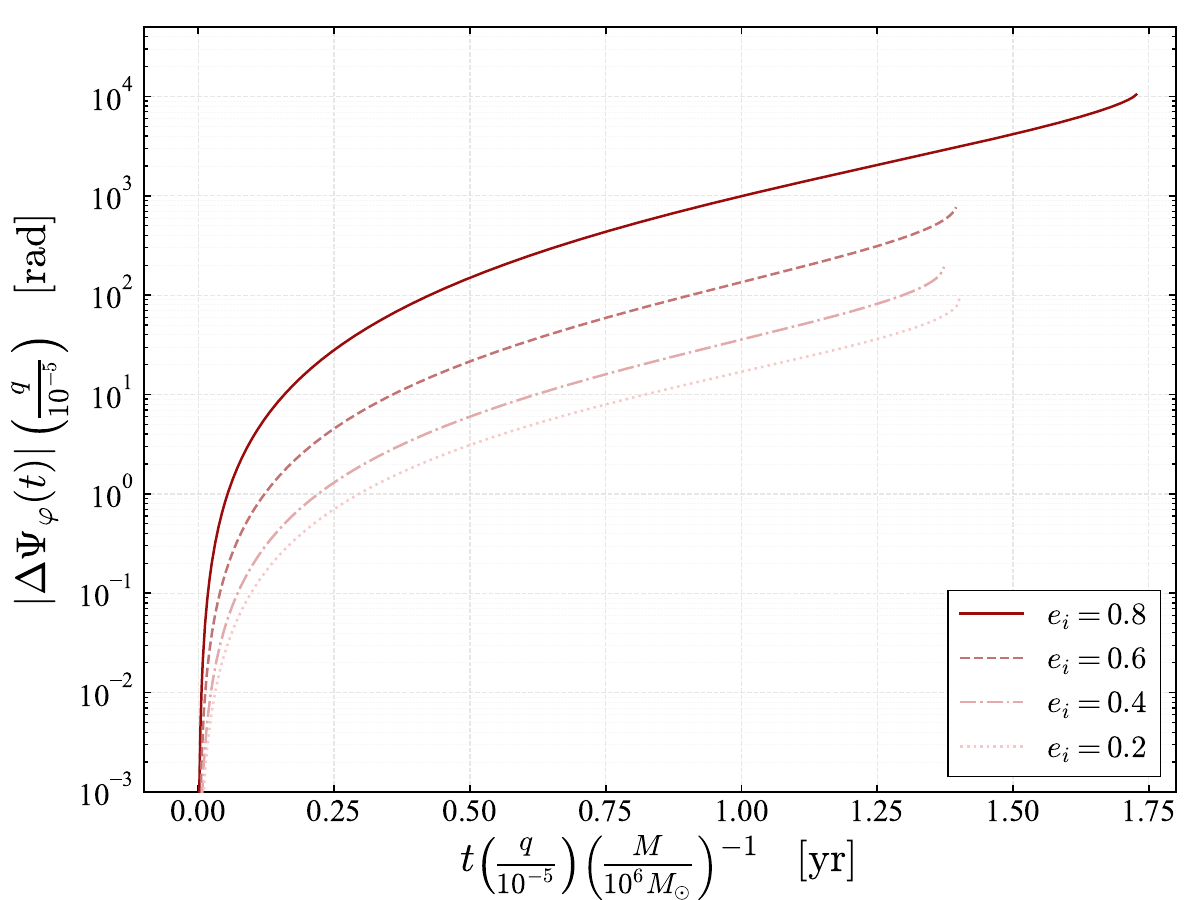}
    \caption{Orbital evolution trajectories in the $(p, e)$ plane (\textit{left panel}) and corresponding dephasing with respect to a vacuum inspiral as a function of time (\textit{right panel}). We consider four systems starting at $p_i = 10$ with initial eccentricities $e_i = 0.8,~0.6,~0.4$ and $0.2$ within a dipolar cloud with the same properties as in Fig.~\ref{fig:Fluxes}. For the $(p, e)$ trajectories, the solid lines include both the GW and scalar field torques, while the dashed lines denote the vacuum inspirals (i.e. GW only).}
    \label{fig:evolution}
\end{figure*}

Most strikingly, a series of resonances that grow in importance with increasing eccentricity and are absent in the circular case can be clearly seen in the scalar-field torque. All these resonances arise from transitions to cloud states with $\ell_f=m_f=0$, occurring at $\sigma_{-1, n}^{n_f}$. The most prominent broad peaks, especially noticeable for $e=0.8$, correspond to resonances with the $n_f=0$ state, in the $n =1, 2, 3, 4$ harmonics (ordered by decreasing $p$). These broad resonances substantially enhance the torques and orbital loss rates, as can be seen by the nearly two orders of magnitude difference between the on-resonance $\dot L^{s,H}$ for $e=0.8$ and for the circular case.
We have verified that these eccentricity-induced resonances persist down to at least $M\mu=0.1$, although the horizon loss rates decrease with decreasing $M\mu$ and the resonances shift to larger orbital separations; see the companion paper~\cite{Xu:2026aic} for a detailed comparison.

In the torque $\dot L^{s,H}$, a clear substructure consisting of a series of narrower peaks can also be seen. These secondary peaks arise from resonances in the subsequent harmonic $n+1$ with $n_f \geq 1$ states, so that close to the peak associated with the $(n,n_f=0)$ resonance, there is always a sequence of narrower peaks arising from $(n+1,n_f \geq 1)$ resonances. Interestingly, the accumulation of resonances occurs to the left of the position where $\dot L^{s,\infty}$ has a sharp jump, which is the opposite behavior of what happens for resonances that occur at larger $p$~\cite{Tomaselli:2025jfo}.

The eccentricity-induced resonances discussed here are absent in a Newtonian treatment, as they arise from $\Omega_{\varphi} \neq \Omega_{r}$ in the strong-field regime.
Close to the LSO, the radial frequency vanishes, $\Omega_r\to0$, while $\Omega_\varphi$ remains finite. Therefore, for the $m=-1$ resonance family, $-\Omega_\varphi+n\Omega_r=\sigma^{n_f}_{-1n}$ can only be satisfied by progressively larger $n$, explaining the accumulation of higher-harmonic resonances near the LSO.

\noindent{\bf{\em Adiabatic inspiral and observational prospects.} }
To assess the impact of eccentricity and resonances in the EMRI dynamics, we model the orbital evolution under the adiabatic approximation.
The trajectory in the $\{p, e\}$ parameter space is driven by the losses in specific energy $\mathcal{E}$ and angular momentum $\mathcal{L}$, including both gravitational and scalar channels. This evolution is governed by
\begin{equation}
\begin{aligned}
&\frac{\dd p}{\dd t} = \frac{1}{H} \left( \frac{\partial \mathcal{L}}{\partial e} \frac{\dd\mathcal{E}}{\dd t} - \frac{\partial \mathcal{E}}{\partial e} \frac{\dd\mathcal{L}}{\dd t} \right), \\
&\frac{\dd e}{\dd t} = \frac{1}{H} \left( \frac{\partial \mathcal{E}}{\partial p} \frac{\dd\mathcal{L}}{\dd t} - \frac{\partial \mathcal{L}}{\partial p} \frac{\dd\mathcal{E}}{\dd t} \right),
\end{aligned}
\end{equation}
where $H$ is the Jacobian of the transformation from $(p, e)$ to $(\mathcal{E}, \mathcal{L})$, given by $H = \frac{\partial \mathcal{E}}{\partial p} \frac{\partial \mathcal{L}}{\partial e} - \frac{\partial \mathcal{L}}{\partial p} \frac{\partial \mathcal{E}}{\partial e}$.

From $p(t)$ and $e(t)$, we compute the orbital frequencies $\Omega_{\varphi,r}(t)$ and quantify the impact of the scalar cloud through the dephasings relative to a vacuum inspiral, 
\begin{equation} \Delta\Psi_{\varphi,r}(t) = \int \Delta\Omega_{\varphi,r} dt . 
\end{equation} 
We find that $|\Delta\Psi_r|$ is typically slightly larger than, but comparable to, $|\Delta\Psi_\varphi|$. Since the dominant gravitational-wave harmonic is typically the $\{m,n\}=\{2,0\}$ mode~\cite{Hughes:2021exa}, whose dephasing is $2\Delta\Psi_\varphi$, we focus on $\Delta\Psi_\varphi$ as a conservative measure of the environmental impact on the observed signal.

We consider systems with initial dimensionless semi-latus rectum $p_i=10$ and eccentricities $e_i \in \{0.2,~0.4,~0.6,~0.8\}$ immersed in a dipolar scalar cloud with the same properties as in Fig.~\ref{fig:Fluxes}. For these four systems, we show the orbital evolution trajectories in the $(p,e)$ plane as well as the corresponding dephasing in Fig.~\ref{fig:evolution}. The infinity loss rates ($\dot E^{s\,,\infty}$, $\dot L^{s\,,\infty}$) have only a modest effect on the orbital evolution, since they are smaller than the horizon contributions across the parameter space considered here (see Fig.~\ref{fig:Fluxes}). We compute the infinity contributions using our fully relativistic framework for all evolutions except the $e_i=0.8$ case, where the large number of radial harmonics required for convergence makes the computation costly; for this case we instead use the Newtonian approximation described in the companion paper~\cite{Xu:2026aic}. We checked that for $e_i=0.6$ this approximation changes the accumulated dephasing by only $\sim4\%$, so we expect its use at $e_i=0.8$ to have a similarly small impact. 
We also verified that the relaxation timescale to reach the steady state at the dominant (broad) $n=1$ resonance, which is expected to scale as $-1/\,{\rm Im}(\omega_{n_f\ell_f}-\omega_{n_i\ell_i})$~\cite{Cardoso:2022Greenhouse}, remains much shorter than the chirp timescale near the resonance for $q\lesssim 10^{-3}$.  This validates the adiabatic treatment of the dominant resonances.

For $e_i=0.8$, there is a clear departure in eccentricity evolution from the vacuum case as shown in the left panel of Fig.~\ref{fig:evolution}. When compared at the same coordinate time, the eccentricity $e(t)$ decreases more slowly in the presence of the scalar cloud. This holds for all eccentricities, but the effect becomes more pronounced as $e_i$ increases, owing to the enhancement of the scalar loss rates (mainly induced by the resonances). For a typical EMRI with $q=10^{-5}$ and $M=10^6\,M_\odot$, the total signal duration from $p_i=10$ to the LSO is approximately one and a half years. Over this period, the accumulated dephasing $\Delta\Psi_\varphi$ relative to vacuum reaches $\sim10^4$~rad for $e_i=0.8$ and $\sim10^3$~rad for $e_i=0.6$, while for smaller eccentricities ($e_i=0.2$) we obtain $\sim 90$~rad, compared to $\sim 70$~rad in the circular limit. Even after half a year of inspiral, the accumulated dephasing for $e_i=0.8$ already exceeds the $e_i=0.2$ case by nearly two orders of magnitude. This indicates that, while eccentricity does little to improve the detectability of the cloud for eccentricities as small as $e_i=0.2$, the improvement becomes substantial at higher eccentricities.

The accumulated dephasing scales approximately linearly with the cloud mass $M_b$. For a lighter cloud with $M_b/M=5\times10^{-3}$ and the same initial semi-latus rectum $p_i=10$, the total dephasing over the entire inspiral is $\sim1.5\times10^3$~rad for $e_i=0.8$ and $\sim 9$~rad for $e_i=0.2$, compared to $\sim 7$~rad in the circular limit. This suggests that the effect should remain detectable for typical LISA signal-to-noise ratios~\cite{Lindblom:2008cm, Khalvati:2024tzz}, even for small cloud masses.\\

Finally we note that, assuming that our results serve as a good proxy for real scalar field clouds in Kerr, the cloud's half-life due to annihilation of scalar particles in GWs scales as $\tau_{\rm cloud}\propto (M/M_b)(M\mu)^{-14}$~\cite{Brito:2017zvb}; in particular for the benchmark configuration with $M\mu=0.2$ and $M_b/M=0.05$,  $\tau_{\rm cloud}\sim 4\times 10^6 {\rm yr}$ for $M=10^6M_\odot$. Although more work is needed to understand how likely it is to find an EMRI in such a cloud, as long as $M\mu\lesssim 0.2$ and $M_b/M$ is not too large, we expect that some clouds can survive enough time in order to be phenomenologically relevant.

\noindent{\bf{\em Discussion.}} 
In this {\it Letter}, we demonstrated that orbital eccentricity plays a key role in the relativistic evolution of EMRIs in scalar-cloud environments. Our most striking finding is the emergence of resonances in the strong-field regime, absent in both circular-orbit models and Newtonian approximations. Resonances substantially enhance the energy and angular-momentum exchange between the EMRI and the scalar cloud, an effect that becomes increasingly pronounced at larger eccentricities. As a result, eccentricity can greatly improve the detectability of scalar clouds, underscoring the importance of eccentricity in both the modeling and interpretation of EMRI signals in boson-cloud environments.

The existence of eccentric EMRIs is well-supported by standard EMRI formation mechanisms via two-body relaxation~\cite{Amaro-Seoane:2007osp,Mancieri:2024sfy, Mancieri:2025cmx}. These mechanisms predict that EMRIs are injected directly into the strong-field region with high eccentricity and a small initial semi-latus rectum ($p\sim 20$~\cite[Eq.~(2)]{ColemanMiller:2005rm}), which could help to bypass resonances encountered at larger separations and avoid long inspirals that could dissipate the cloud before the system enters the LISA band~\cite{Tomaselli:2024bdd,Boskovic:2024fga,Boskovic:2025ixx}. 

A natural extension of our work would be to consider a Kerr MBH. Although superradiant growth may spin the BH down to small values, the tri-periodic nature of generic Kerr orbits could still enrich further the resonant signatures identified here.

\noindent{\bf{\em Acknowledgements.}} 
We are grateful to Seth Hopper for kindly sharing his code developed in Ref.~\cite{Hopper:2010uv} with us. We are grateful to Conor Dyson, Robrecht Keijzer and Thomas Spieksma for insightful discussions and comments on the draft of this manuscript. We acknowledge financial support provided by FCT – Fundação para a Ciência e a Tecnologia, I.P., through the ERC-Portugal program Project ``GravNewFields''. We also thank the Fundação para a Ciência e Tecnologia (FCT), Portugal, for the financial support to the Center for Astrophysics and Gravitation (CENTRA/IST/ULisboa) through grant No.~\href{https://doi.org/10.54499/UID/PRR/00099/2025}{UID/PRR/00099/2025} and grant No.~\href{https://doi.org/10.54499/UID/00099/2025}{UID/00099/2025}.
Q.~X. gratefully acknowledges support from FCT through Grant No.~\href{https://doi.org/10.54499/2025.01546.BD}{2025.01546.BD}.
R.~V. gratefully acknowledges the support of the Dutch Research Council (NWO) through an Open Competition Domain Science-M grant, project number OCENW.M.21.375. 
C.~Y. acknowledges the financial support provided under the European Union's H2020 ERC Advanced Grant ``Black holes: gravitational engines of discovery'' grant agreement no.~Gravitas--101052587. C.~Y. acknowledges support from the Villum Investigator program supported by the VILLUM Foundation (grant no.~VIL37766) and the DNRF Chair program (grant no.~DNRF162) by the Danish National Research Foundation.

 Views and opinions expressed are however those of the author only and do not necessarily reflect those of the European Union or the European Research Council. Neither the European Union nor the granting authority can be held responsible for them.

\bibliography{./References}
%
\end{document}